\documentclass[a4paper, 11pt]{article}

\usepackage{amsmath,amssymb,amsfonts}
\usepackage{graphicx}
\usepackage{color}
\usepackage{appendix}
\usepackage{authblk}
\usepackage{amsthm}
\usepackage{tikz}

\newcommand{\nn}{\nonumber}
\newcommand{\bn}{\begin{enumerate}}
	\newcommand{\en}{\end{enumerate}}

\newcommand{\bm}{\boldmath}
\def\mf{\mathbf}

\def\bra#1{\mathinner{\langle{#1}|}}
\def\ket#1{\mathinner{|{#1}\rangle}}

\def\d{\delta}

\def\r{\rho}

\def\<{\langle}
\def\>{\rangle}

\def\tr{{\rm tr}}

\def\jmath{{j}}

\begin{document}

\title{Quantum Computing for Molecular Vibronic Spectra and Gaussian Boson Sampling}

\author{Seungbeom Chin}
\author{Joonsuk Huh\thanks{joonsukhuh@skku.edu}}
\affil{Department of Chemistry, Sungkyunkwan University, Suwon 16419, Korea}
\maketitle
\begin{abstract}
    Boson sampling (BS) is a multimode linear optical problem that is expected to be intractable on classical computers. It was recently suggested that molecular vibronic spectroscopy (MVS) is computationally as complex as BS. In this review, we discuss the correspondence relation between BS and MVS and briefly introduce the experimental demonstrations of the molecular spectroscopic process using quantum devices. The similarity of the two theories results in another BS setup, which is called ``vibronic BS''.  The hierarchical structure of vibronic BS, which includes the original BS and other Gaussian BS, is also explained.
\end{abstract}

\section{Introduction}

A multimode quantum optical system called  ``Boson Sampling (BS)'' \cite{aaronson2011computational} is considered to demonstrate quantum speedup with fewer implements than the universal quantum computer. The original BS is a strong candidate for a counterexample to the extended Church-Turing (ECT) thesis, which states that everything efficiently computable with actual devices are efficiently solvable by a Turing machine.
In the BS system, the input state is a set of non-interacting identical single photons, which are sent through a linear optical network and result in another photon distribution. This process is not classically accessible at a sufficiently large scale.

Instead of non-interacting single photon sources, Gaussian states recently have been suggested as input states for BS \cite{Lund2014, rahimi2015}. An interesting application of Gaussian BS is that it can simulate molecular spectroscopy \cite{huh2015boson}. In the process of molecular vibronic spectroscopy (MVS), we consider the vibrational transitions between two electronic states of a molecule. A molecule undergoes a structural deformation through vibronic transition. An experiment was designed in \cite{huh2015boson} to simulate MVS with BS, since the Frank-Condon profiles (FCPs, defined as the probability distribution regarding a given transition) correspond to the transition amplitudes of the Fock state BS. Actual experiments were performed recently with trapped ions \cite{shen2017quantum} and photonics \cite{clements2017experimental}.

On the other hand, this intimate relation between BS and MVS yielded another version of BS, which is named ``vibronic BS'' \cite{huh2017vibronic}. With the most general Gaussian states as an input and twice as many modes as the original setup, one can find a hierarchical structure that embraces every instance of the known Gaussian BS.

In this review, we discuss the correspondence between BS and MVS, and briefly introduce the experimental demonstrations of the molecular spectroscopic process using the correspondence. Then the concept of vibronic BS and its hierarchical structure \cite{huh2017vibronic} is explained.

\section{Boson sampling and vibronic transition}
\indent 

We first summarize the theoretical background for the whole discussion, focusing on the definition of Gaussian BS and molecular vibronic transition. 

\subsection{Boson Sampling and the Gaussian version}
\indent

The general form of $P_{\hat{\Pi}}$, the probability of a post-selected measurement $\hat{\Pi}$ for sampling Bosons, is given by
\begin{align}
P_{\hat{\Pi}}=\tr\left[\hat{O}\hat{\rho}_{\mathrm{in}}\hat{O}^{\dagger}\hat{\Pi}\right],
\end{align}
where $\hat{O}$ is a transition operator that represents the given quantum optical network, and $\hat{\rho}_{\mathrm{in}}$ is an input state.
There exist various boson sampling setups that are determined by $(\hat{\rho}_{\mathrm{in}}, \hat{O}, \hat{\Pi})$, e.g., 
\begin{itemize}
	\item The original boson sampling~\cite{aaronson2011computational}:  $\hat{\Pi}=\ket{\mathbf{m}}\bra{\mathbf{m}}$, $\hat{\rho}_{\mathrm{in}}=\ket{\mf{n}}\bra{\mf{n}}$ (here $\mf{m}$ and $\mf{n}$ are photon distribution vectors of $N$ photons in $M$ ($\gg N$) modes, and the components of ($\mf{m},\mf{n}$) are not larger than 1, therefore $\hat{\prod}$ is a photon number projector), and $\hat{O}=\hat{R}_{U}$, a beam splitting operator with a unitary matrix $U$. The exact definition of $\hat{R}_U$ is given in Appendix \ref{operators}, which shows that  $\hat{O}$ does not change the total photon number $N$.
	\item Gaussian boson sampling \cite{rahimi2015}: $\hat{\Pi}=\ket{\mathbf{m}}\bra{\mathbf{m}}$, $\hat{\rho}_{\mathrm{in}}$ are products of Gaussian modes ($e.g.$, squeezed vacuum states or thermal states), and $\hat{O}=\hat{R}_{U}$,
\end{itemize}  
The most general expression for a Gaussian state $\hat{\rho}_{\mathrm{G}}$ is obtained from a thermal state $\hat{\rho}_{\mathrm{th}}$ and a Gaussian operator $\hat{O}_{\mathrm{G}}$~\cite{weedbrook2012gaussian, adesso2014}, i.e.   $\hat{\rho}_{\mathrm{G}}=\hat{O}_{\mathrm{G}}\hat{\rho}_{\mathrm{th}} \hat{O}_{\mathrm{G}}^{\dagger}$. 
The action of $\hat{O}_{\mathrm{G}}$ on the Boson creation operator, $\hat{\mathbf{a}}^{'\dagger} = \hat{O}_{\mathrm{G}}^{\dagger}\hat{\mathbf{a}}^{\dagger} \hat{O}_{\mathrm{G}}$ is generally expressed  as 
\begin{equation}
\hat{\mathbf{a}}^{'\dagger} = \hat{O}_{\mathrm{G}}^{\dagger}\hat{\mathbf{a}}^{\dagger} \hat{O}_{\mathrm{G}} = X\hat{\mathbf{a}}+Y\hat{\mathbf{a}}^{\dagger}+\mathbf{z} \ ,
\label{eq:genBT}
\end{equation}  
where  $X$ and $Y$ ($M\times M$ matrices) satisfy $XX^{\dagger}-YY^{\dagger}=I$ and $XY^{\mathrm{t}}=YX^{\mathrm{t}}$, and 
$\hat{\mathbf{a}}^{'\dagger}$ and $\hat{\mathbf{a}}^{\dagger}$ are the $M$-dimensional boson creation operator column vector of output and input states, respectively (see Appendix \ref{operators}).                               
Since $\hat{O}_{\mathrm{G}}$ can be decomposed into quantum optical operators as
\begin{align}
\hat{O}_{\mathrm{G}}=\hat{D}_{\mathbf{z}^{*}}\hat{R}_{U_{\mathrm{L}}}\hat{S}_{\Sigma}\hat{R}_{U_{\mathrm{R}}^{\dagger}},
\end{align}
 ( $\hat{D}_{\mathbf{z}^{*}}$ is the displacement operator, $\hat{R}_U$ is the rotation operator, and $\hat{S}_{\Sigma}$ is the squeezing operator. For the detailed definitions and relations for these operators, see Appendix \ref{operators}),
 $X$ and $Y$ in Eq.~\ref{eq:genBT} are identified by singular value decomposition with two unitary matrices $U_{\mathrm{L}}$ and $U_{\mathrm{R}}$ and a real diagonal matrix $\Sigma$ as follows:  $X=U_{\mathrm{L}}\sinh(\Sigma)U_{\mathrm{R}}^{\mathrm{t}}$ and $Y=U_{\mathrm{L}}\cosh(\Sigma)U_{\mathrm{R}}^{\dagger}$. 

\subsection{Vibronic transition and Franck-Condon Profile}
\indent

It was first pointed out in \cite{huh2015boson} that a system of $N$ photons in $M$ modes is isomorphic to that of $N$ phonons in $M$ vibrational modes. To understand the relation clearly, we review the concept of vibrational transition of molecules and Franck-Condon Profile (FCP).

The total molecular wavefunction of the nuclei at the position vector ${\bf \hat{R}} = (\hat{R}_1, \hat{R}_2, ...) $ and electrons at ${\bf \hat{r}} = (\hat{r}_1,\hat{r}_2,...)$ is simplified as $  \psi( {{\mathbf{\hat{r}}},{\mathbf{\hat{R}}}} )\phi ( {\mathbf{\hat{R}}})$ under the Born-Oppenheimer approximation (note that the electronic wavefunction $\psi ( {{\mathbf{\hat{r}}},{\mathbf{\hat{R}}}})$ depends on the nuclear coordinates $\bf \hat{R}$). 
Therefore, for transitions between two electronic levels $\ket{g}$ (the ground state) and $\ket{e}$ (the excited state), the molecular Hamiltonian $\cal{H}_{\rm mol}$ is approximately given by
\begin{align}
{\cal{H}_{{\text{mol}}}}
 &= \left| g \right\rangle \left\langle g \right| \otimes  \Big( \sum\nolimits_k \hbar{{\omega _k}\hat{a}_k^\dag {\hat{a}_k}} \Big)  + \left| e \right\rangle \left\langle e \right| \otimes \Big( \sum_k \hbar \omega'_k\hat{a}'^\dagger_k\hat{a}'_k + \hbar \omega_{\textrm{ad}} \Big) \nn \\
&\equiv  \left| g \right\rangle \left\langle g \right| \otimes {H_g} + \left| e \right\rangle \left\langle e \right| \otimes {H_e},
\end{align}
where $H_g$ is the nuclear Hamiltonian for the ground state $|g\>$ and $H_e$ for the excited state $|e\>$ ($\omega_k$ and $\omega_{k}'$ are the angular frequencies for $k$-th mode of the initial and final electronic states, respectively.). The electronic adiabatic transition frequency is denoted as $\omega_{\rm ad}$.   

The normal coordinate of initial ($\mathbf{Q}$) and final ($\mathbf{Q'}$) states have the Duschinsky relation~\cite{duschinsky1937importance}, $\mathbf{Q'=UQ+d}$. $\mathbf{U}$ and  $\mathbf{d}$ are the Duschinsky rotation matrix and displacement vector, respectively. 
The corresponding Bogoliubov transformation for the vibronic transition is given  by the Doktorov operator $\hat{U}_{\mathrm{D}}$ \cite{doktorov1976}, which is expressed as 
\begin{align}
\mathbf{\hat{a}}^{'\dagger}=
\hat{U}_{\mathrm{D}}^\dagger \mathbf{\hat{a}}^{\dagger} \hat{U}_{\mathrm{D}}
= \frac{1}{2}\left(J-(J^{\mathrm{t}})^{-1}\right)\mathbf{\hat{a}}
+\frac{1}{2}\left(J+(J^{\mathrm{t}})^{-1}\right)\mathbf{\hat{a}}^{\dagger}+\frac{1}{\sqrt{2}}\boldsymbol{\delta} \, ,
\label{eq:duschinskya}
\end{align}
where $J$ and $\boldsymbol{\delta}$ are defined as 
\begin{align}
&J_{ij}=\Omega_i'U_{ij}\Omega_j, \quad
\boldsymbol{\delta}_i=\hbar^{-\tfrac{1}{2}}\Omega'_i d_i,
\label{eq:parameters}
\end{align}
where $\mf{\Omega}'$ and $\mf{\Omega}$ are the vectors for square roots of angular frequencies of the final and initial states, i.e., $\mf{\Omega'}=(\sqrt{\omega_{1}'},\ldots,\sqrt{\omega_{N}'})$ and $\mf{\Omega}=(\sqrt{\omega_{1}}^{-1},\ldots,\sqrt{\omega_{N}}^{-1})$.  It is  straightforward to see that Eq. \eqref{eq:duschinskya} is a special case of Eq. \eqref{eq:genBT} with
\begin{align}\hat{O}_\textrm{G} = \hat{U}_\textrm{D}, \quad 
X=\frac{1}{2}\left(J-(J^{\mathrm{t}})^{-1}\right), \quad Y= \frac{1}{2}\left(J+(J^{\mathrm{t}})^{-1}\right),\quad \mf{z} = \frac{1}{\sqrt{2}}\boldsymbol{\delta}.
\label{fcog}
\end{align}
The Doktorov operator $\hat{U}_\textrm{D}$ is expressed with optical operators as \cite{doktorov1976}
\begin{align}
\hat{U}_\textrm{D}
=\hat{D}_{\mathbf{\d}/\sqrt{2}}\hat{S}^\dagger_{\Omega'}\hat{R}_{U}\hat{S}_{\Omega},
\label{uddecomp}
\end{align}
where $\hat{D}$, $\hat{R}$, and $\hat{S}$ are the displacement, rotation, squeezing operators (see Appendix \ref{operators}).
FCP is an essential quantity that is proportional to the molecular absorption lineshape, 
\begin{align}\label{FCP_thermal}
\mathrm{FCP}\left( \omega_{\mathrm{v}}  \right) = \sum\limits_{\mathbf{m}} {{{| {\langle {\mathbf{m}} |\hat{U}_{\mathrm{D}}| {\mathbf{n}} \rangle } |}^2}{P_{{\text{in}}}}\left( {\mathbf{n}} \right){\delta _{\omega_{\mathrm{v}}} (\Delta) }}.
\end{align}
Here $|\< \mathbf{m}|\hat{U}_{\mathrm{D}}| {\mathbf{n}} \> |^2$ is the Franck-Condon factor, ${{\delta _{\omega_{\mathrm{v}}} (\Delta) }}$ with $\Delta  = \omega_{\mathrm{v}}  - {\mathbf{m}} \cdot {\bm \omega} ' + {\mathbf{n}} \cdot {\bm \omega} $ imposes the energy conservation, and  $P_{\rm in}({\bf n})$ denotes the initial distribution of the phonon mode.

The relation of operators given by Eq. \eqref{uddecomp} renders a quantum optical network with multimodes to simulate the FCP efficiently, as will be explained in more detail in the next section.  

\section{Simulating MVS with BS}
\indent

Since the vibronic transition is reexpressed as a special case of the Gaussian state operation as can be seen in Eq. \eqref{uddecomp}, we can set up a BS system so that it can simulate MVS. 
By taking the optical operators Eq. \eqref{uddecomp} subsequently to the vacuum state and  measuring the final state in the Fock basis as in Eq. \eqref{FCP_thermal}, one can
simulate the FCP with quantum devices.
The brief summary of the simulation protocol, with trapped-ion  as a concrete example \cite{shen2017quantum}, is as follows: (i) initialize the ion in the ground state, (ii) apply the Doktorove operation \eqref{uddecomp} to the ion, (iii) detect the vibronic spectrum of final state with the projection measurements. 

In this section we introduce such BS simulations with various devices.
Even though the standard BS requires less optical operations than the Gaussian BS, the difficulties in preparing the initial states render the former more challenging in practical optical systems. Indeed, it is hard to create single photon states (the
original boson sampling) and squeezed coherent states (the molecular simulation) in realistic laboratories.
 The candidates suggested for scalable BS are trapped ions \cite{lau2012,shen2014} and superconducting circuits \cite{peropadre2016}. 
The photoelectron spectra of SO$_2$ and SO$^-_2$ were demonstrated with a trapped ion device by Yangchao et al. \cite{shen2017quantum}. Recently, Clements et al. \cite{clements2017experimental} attempted to simulate the absorption spectrum of tropolone  (C$_7$H$_{6}$O$_2$) with a photonic setup.

	

\subsection{Simulation with trapped ions}
\indent

The first experimental
demonstration of the MVS of SO$_2$ with 
trapped-ions was reported in \cite{shen2017quantum}.
For a reliable Gaussian BS, the experimental technology for phonons to be operated with displacement, squeezing, and rotation is developed. The operators that can simulate the optical systems are installed through Raman laser beams, resulting in a Bogoliubov transformation. As a result, the photoelectron spectroscopies of SO$_2$ and SO$_2^-$ is reproduced with the trapped-ion simulator.

\subsection{Simulation with optics}
\indent

It was demonstrated in \cite{clements2017experimental} that an approximated pattern of the molecular vibronic spectra can be simulated with imperfect quantum simulators. The partial vibronic spectrum of tropolone is experimentally simulated to prove the approximation claim. The result shows that there exists the efficient error bound on the estimations of the spectra and such simulations can
surpass a classicality criterion.

\subsection{Simulation with superconducting circuits}
\indent

Spectroscopy is an important technique for interpreting quantum systems with electromagnetic radiation. However, technical limitations restrict the practical application of molecular spectroscopy. It was demonstrated in \cite{hu2017} that a superconducting quantum simulator can perform a simple one-dimensional computation.

\section{Vibronic boson sampling}\label{VBS}
\indent

In this section, we introduce a hierarchical  structure \cite{huh2017vibronic} that encompasses all the known types of BS, including 
scattershot BS~\cite{Lund2014}, and Gaussian BS~\cite{rahimi2015}. We also show that it is possible to incorporate the initial thermal Gaussian correlation with the Gaussian BS with no correlation. 
It is achieved by doubling the number of modes of Scattershot BS and the size of the photon network.

We first present the definition of vibronic BS, which has an initial state that has the purification form of mixed thermal states with ancillar modes 
 i.e., 
$\ket{\mathbf{0}
	(\boldsymbol{\beta})}=\sum_{\mathbf{n}=\mathbf{0}}^{\boldsymbol{\infty}}\sqrt{\bra{\mathbf{n}}\hat{\rho}_{\mathrm{th
	}}\ket{\mathbf{n}}}\ket{\mathbf{n}}\otimes\ket{\mathbf{n}}_{\mathrm{B}}$,
where $\hat{\r}_{\mathrm{th}}$ is the thermal state and $|\mf{n}\>_{\mathrm{B}}$ is the Fock state of the ancillary system with $M$ modes. Note that $\tr_{\mathrm{B}}[\ket{\mathbf{0}(\boldsymbol{\beta})}\bra{\mathbf{0}
	(\boldsymbol{\beta})}] = \hat{\rho}_{\mathrm{th}}$ (tr$_\textrm{B}$ is the partial trace over the ancillary system).
The pure state $\ket{\mathbf{0}(\boldsymbol{\beta})}$ can be rewritten as
\begin{align}
\ket{\mathbf{0}(\boldsymbol{\beta})}=\hat{V}
(\boldsymbol{\beta})\ket{\mathbf{0}}\otimes\ket{\mathbf{0}}_{\mathrm{B}},
\end{align} where the operator $\hat{V}(\boldsymbol{\beta})$ is a tensor product of two-mode squeezing operators, i.e., 
$\hat{V}(\boldsymbol{\beta})=\bigotimes_{k=1}^{M}\exp(\theta_{k}
(\hat{a}_{k}^{\dagger}\hat{b}_{k}^{\dagger}-\hat{a}_{k}\hat{b}_{k})/2)$,
where 
$\tanh(\theta_{k}/2)=\mathrm{e}^{-\beta_{k}\hbar\omega_{k}/2}=\sqrt{\bar{n}_{k}/(\bar{n}_{k}+1)}$, 
$\hat{b}_{k}$ ($\hat{b}_{k}^{\dagger}$) is the annihilation (creation) operator for the ancillary system, and  ${{\bar n}_k} = 1/({e^{\beta_k \hbar \omega_k }} - 1)$ is the average number of photons in the $k$-th mode.
The linear optical operation $\hat{O}_\mathrm{G}$ acts on both the original and ancillary Hilbert space.
As a result, the BS with initial sampling of the Fock state is translated to a problem only with the post-selection measurement, i.e., 

\begin{align}
P(\mathbf{m},\mathbf{n})=\tr\Big[\hat{O}_{\mathrm{G}} \ \ket{\mathbf{0}
	(\boldsymbol{\beta})}\bra{\mathbf{0}
	(\boldsymbol{\beta})}\hat{O}_{\mathrm{G}}^{\dagger}\ket{\mathbf{m}}\bra{\mathbf{m}}\otimes\ket{\mathbf{n}}_{\mathrm{B}}\mathrm{_B{\bra{\mathbf{n}}}{}} \Big],
\label{eq:proboperator}
\end{align}
where the trace is on the overall Hilbert space. 
When the Gaussian operator is the rotation operator, i.e. $\hat{O}_{\mathrm{G}}=\hat{R}_{U}$,
Eq.~\eqref{eq:proboperator} is reduced to the scattershot BS \cite{Lund2014}. Therefore, we can state that the scattershot BS is a special case of vibronic BS.

In the following, we show that a Bogoliubov transformation can eliminate the mode correlation from the two-mode squeezing operator $\hat{V}$ \cite{huh2017vibronic}.
With the definition
\begin{equation}
\hat{U}(\boldsymbol{\beta}) \equiv \hat{O}_{\mathrm{G}} \ \hat{V}(\boldsymbol{\beta}),
\end{equation}
the Bogoliubov transformation by $\hat{U}(\boldsymbol{\beta})$ on the collective creation operator vector $\mathbf{c}^{\dagger}$ ($\equiv (\mathbf{a}^{\dagger},\mathbf{b}^{\dagger})^{\mathrm{t}}$) of the extended space is given by 
\begin{equation}
\hat{\mathbf{c}}^{'\dagger}=\hat{U}(\boldsymbol{\beta})^{\dagger}\hat{\mathbf{c}}^{\dagger}\hat{U}(\boldsymbol{\beta})=\mathcal{X}\hat{\mathbf{c}}+\mathcal{Y}\hat{\mathbf{c}}^{\dagger}+\boldsymbol{\gamma},
\label{bogolextended}
\end{equation}
which is an extension of Eq. \eqref{eq:genBT}.
The parameters $\mathcal{X}$, $\mathcal{Y}$, and $\gamma$ are written as 
\begin{align}
\mathcal{X}=
\begin{pmatrix}
XF & YG\\
G & \mathrm{diag}(\mathbf{0})
\end{pmatrix}, \quad
\mathcal{Y}=
\begin{pmatrix}
YF & XG\\
\mathrm{diag}(\mathbf{0}) & F
\end{pmatrix}, \quad 
\boldsymbol{\gamma}=
\begin{pmatrix}
\mathbf{z}\\
\mathbf{0}
\end{pmatrix},
\end{align}
where 
$F\equiv \mathrm{diag}(\sqrt{\bar{n}_{1}+1},\ldots,\sqrt{\bar{n}_{M}+1})$ and  
$G\equiv \mathrm{diag}(\sqrt{\bar{n}_{1}},\ldots,\sqrt{\bar{n}_{M}})$. 

Using Eq. \eqref{bogolextended}, one can convert vibronic BS into Gaussian BS with squeezed coherence states ($\gamma \neq \bf{0}$) or squeezed vacuum states ($\gamma =\bf{0}$) as inputs. 
It is achieved by the singular valued decompostion of the matrices  
$\mathcal{X}=\mathcal{C}_{\mathrm{L}}\sinh(\mathcal{S})\mathcal{C}_{\mathrm{R}}^{\mathrm{t}}$ and $\mathcal{Y}=\mathcal{C}_{\mathrm{L}}\cosh(\mathcal{S})\mathcal{C}_{\mathrm{R}}^{\dagger}$ \ ,
where $\mathcal{C}_{\mathrm{L}}$ and $\mathcal{C}_{\mathrm{R}}$ are the unitary matrices and $\mathcal{S}=\textrm{diag}(s_{1},\ldots,s_{2M})$ is a real diagonal matrix that includes the squeezing parameters. 

The extended unitary operator is decomposed as
\begin{align}
\hat{U}(\boldsymbol{\beta})=\hat{D}_{\boldsymbol{\gamma}^{*}}\hat{R}_{\mathcal{C}_{\mathrm{L}}}\hat{S}_{\mathcal{S}}\hat{R}_{\mathcal{C}_{\mathrm{R}}^{\dagger}}=
\hat{R}_{\mathcal{C}_{\mathrm{L}}}\hat{S}_{\mathcal{S}}\hat{R}_{\mathcal{C}_{\mathrm{R}}^{\dagger}}\hat{D}_{\boldsymbol{\gamma}'},
\label{eq:decomposition}
\end{align}
where the displacement parameter vector  $\boldsymbol{\gamma}'=\boldsymbol{\gamma}'_{\mathrm{R}}+\mathrm{i}\boldsymbol{\gamma}'_{\mathrm{I}}$ is given by, 
\begin{align}
\begin{pmatrix}
\boldsymbol{\gamma}'_{\mathrm{R}}\\
\boldsymbol{\gamma}'_{\mathrm{I}}
\end{pmatrix}
=
\begin{pmatrix}
\mathcal{X}_{\mathrm{R}}+\mathcal{Y}_{\mathrm{R}} & -\mathcal{X}_{\mathrm{I}}+\mathcal{Y}_{\mathrm{I}}\\
\mathcal{X}_{\mathrm{I}}+\mathcal{Y}_{\mathrm{I}} & \mathcal{X}_{\mathrm{R}}-\mathcal{Y}_{\mathrm{R}}
\end{pmatrix}^{-1}
\begin{pmatrix}
\boldsymbol{\gamma}_{\mathrm{R}}\\
\boldsymbol{\gamma}_{\mathrm{I}}
\end{pmatrix} ,
\label{eq:leqrealimag}
\end{align}
with $\boldsymbol{\gamma}=\boldsymbol{\gamma}_{\mathrm{R}}+\mathrm{i}\boldsymbol{\gamma}_{\mathrm{I}}$, $\mathcal{X}=\mathcal{X}_{\mathrm{R}}+\mathrm{i}\mathcal{X}_{\mathrm{I}}$ and $\mathcal{Y}=\mathcal{Y}_{\mathrm{R}}+\mathrm{i}\mathcal{Y}_{\mathrm{I}}$.
One can implement the operator decomposed as the second equality  of Eq. \eqref{eq:decomposition} by preparing the $2M$-mode squeezed coherent states 
\begin{equation}
\hat{S}_{\mathcal{S}}\hat{R}_{\mathcal{C}_{\mathrm{R}}^{\dagger}}\hat{D}_{\boldsymbol{\gamma}'}\ket{\mathbf{0}} \equiv \hat{S}_{\mathcal{S}}\vert \boldsymbol{\gamma}''\rangle 
=\bigotimes_{k=1}^{2M}\hat{S}_{s_{k}}\vert \gamma_{k}''\rangle\ ,
\label{eq:twomsinglemode}
\end{equation}
which are injected to the quantum optical device, and then taking the linear optical operation ($\hat{R}_{\mathcal{C}_{\mathrm{L}}}$).
The vector $\boldsymbol{\gamma}''$ is obtained by a rotation of $\boldsymbol{\gamma}'$, i.e. $\boldsymbol{\gamma}''=C_{\mathrm{R}}^{\mathrm{t}}\boldsymbol{\gamma}'$.  
Hence, we see that the $2M$-mode squeezed coherent states replace the correlated squeezed thermal state $\hat{S}\hat{R}\rho_{\mathrm{th}}\hat{R}^{\dagger}\hat{S}^{\dagger}$.

The hierarchical structure of Gaussican BS is described as follows:
\begin{align}
&\textrm{Gaussian BS (M modes, T=0)} =\textrm{vibronic BS (M modes, T=0)} \nn \\
& \subseteq \textrm{Gaussian BS (M modes, T$\neq$0)} 
 \subseteq \textrm{vibronic BS (M modes, T$\neq$0)} \nn \\
& \subseteq \textrm{Gaussian BS (2M modes, T=0)}  \subseteq \textrm{vibronic BS (M modes, T=0)} \nn
\end{align}


		

The hierarchy of Gaussian BS is arrayed along the number of modes $\#$(mode) and temperature T. When $\#$(mode)  and T are the same, vibronic BS is always a more general case of Gaussian BS. The only exception is when T=0, for which the two types of BS are equivalent \cite{huh2015boson}. The Gaussian BS with $\#$(mode)=$M$ and T$\neq 0$ is the case analyzed in \cite{rahimi2015}. In addtion, the hierarchical relation ``original BS $\subseteq$ scattershot BS $\subseteq$ vibronic BS'' holds in the extended Hilbert space, which is confirmed from the fact that the scattershot BS with post-selected measurement becomes the original BS \cite{Lund2014}.

\section{Conclusions}
\indent

In this review, we explained how the isomorphism between BS and MVS presents a distinctive viewpoint on some aspects of quantum computation problems. The mathematical relation of Gaussian and vibronic BS is exploited in experiments to simulate MVS with non-optical BS systems, and to construct the hierarchical structure of vibronic BS that includes other known BS setups. We expect that the complexity analysis of vibronic BS would provide a new insight to understand the extended Church-Turing thesis with BS. Finally, we comment on the possible extension of vibronic BS to go beyond the harmonic picture of the oscillators and the Condon approximation for the transition dipole moment. For the former, we would be able to incorporate the simple anharmonicity by a Morse oscillator \cite{Iachello:1998,Santiago2017}. For the latter, one can approximate the non-Condon operator \cite{Huh2012} as a unitary operator so that the current vibronic boson sampling setup can be simply used. Currently, we are working in these directions for the quantum simulation of vibronic problems.        

\section{Acknowledgements}

This work was supported by Basic Science Research Program through the National Research Foundation of Korea (NRF) funded by the Ministry of Education, Science and Technology (NRF-2015R1A6A3A04059773 and 2017R1A4A1015770). 

\appendix
\section{Quantum optical operators}\label{operators}

We denote the tranformation of boson operator vector $\hat{\mathbf{x}}$ as
$\hat{A} \ \hat{\mathbf{x}} \ \hat{B} \equiv (\hat{A}\hat{x}_{1}\hat{B},\ldots,\hat{A}\hat{x}_{M}\hat{B})^{\mathrm{t}}$.
The optical unitary operators are defined as
\begin{align}
&\hat{R}_{U}=\exp((\hat{\mathbf{a}}^{\dagger})^{\mathrm{t}}\ln U^{*}\hat{\mathbf{a}}) \qquad(\textrm{rotation}), \nn \\
&\hat{D}_{\boldsymbol{\alpha}}=\exp(\boldsymbol{\alpha}^{\mathrm{t}} \hat{\mathbf{a}}^{\dagger}-\boldsymbol{\alpha}^{\dagger}\hat{\mathbf{a}}) \qquad(\textrm{displacement}), \nn \\
&\hat{S}_{\Sigma}=\exp(((\hat{\mathbf{a}}^{\dagger})^{\mathrm{t}}\Sigma\hat{\mathbf{a}}^{\dagger}-\hat{\mathbf{a}}^{\mathrm{t}}\Sigma\hat{\mathbf{a}})/2) \qquad(\textrm{squeezing}),
\end{align}
where $\boldsymbol{\alpha}$ is a coherent state phase vector in $d=M$, $\Sigma$ is a diagonal matrix, and $U$ is an $M\times M$ unitary matrix. 
The optical operators act on $\mathbf{\hat{a}}^{\dagger}$ as
\begin{align}
&\hat{R}_{U}^{\dagger}\hat{\mathbf{a}}^{\dagger}
\hat{R}_{U}
= U\hat{\mathbf{a}}^{\dagger}, \nn \\
&\hat{D}_{\boldsymbol{\alpha}}^{\dagger}\mathbf{\hat{a}}^{\dagger}\hat{D}_{\boldsymbol{\alpha}}=\mathbf{\hat{a}}^{\dagger}+\boldsymbol{\alpha}^{*}, 
\nn
\\
&\hat{S}_{\Sigma}^{\dagger}\mathbf{\hat{a}}^{\dagger}\hat{S}_{\Sigma}=
\sinh(\Sigma)\mathbf{\hat{a}}
+\cosh(\Sigma)\mathbf{\hat{a}}^{\dagger},
\end{align}
 and the squeezing operator $\hat{V}(\boldsymbol{\beta})=\bigotimes_{k=1}^{M}\exp(\theta_{k}
(\hat{a}_{k}^{\dagger}\hat{b}_{k}^{\dagger}-\hat{a}_{k}\hat{b}_{k})/2)$ acts as
\begin{align}
\hat V{(\bm{\beta} )^\dag }{{\hat a}_k}\hat V(\bm{\beta} ) & = \cosh ({\theta _k}/2) \ {{\hat a}_k} + \sinh ({\theta _k}/2) \ \hat b_k^\dag  \hfill , \nn \\
\hat V{(\bm{\beta} )^\dag }{{\hat b}_k}\hat V(\bm{\beta} ) & = \sinh (\theta_k/2) \ \hat a_k^\dag  + \cosh ({\theta _k}/2) \ {{\hat b}_k} \hfill .
\end{align}

\bibliographystyle{iopart-num}
\providecommand{\newblock}{}

\end{document}